\documentclass[a4paper,preprint]{revtex4}
\usepackage{epsf}
\usepackage[dvips]{graphicx}
\usepackage{graphics}

\begin{document}
\setcounter{page}{1}
\title{Numerical Evidence for Stretched Exponential Relaxations in the Kardar-Parisi-Zhang Equation}
\author{Eytan Katzav}
\email{eytak@post.tau.ac.il}
\author{Moshe Schwartz}
\email{mosh@tarazan.tau.ac.il} \affiliation {School of Physics and Astronomy, Raymond and Beverly Sackler
Faculty of Exact Sciences, Tel Aviv University, Tel Aviv 69978, Israel}

\begin{abstract}
We present results from extensive numerical integration of the KPZ
equation in $1 + 1$ dimensions aimed to check the long-time
behavior of the dynamical structure factor of that system. Over a
number of decades in the size of the structure factor we confirm
scaling and stretched exponential decay. We also give an analytic
expression that yields a very good approximation to the numerical
data. Our result clearly favors stretched exponential decay over
recent results claiming to yield the exact time dependent
structure factor of the $1+1$ dimensional KPZ system. We suggest a
possible solution to that contradiction.
\end{abstract}

\maketitle

Many interesting dynamical phenomena in condensed matter physics are described in terms of non-linear field
equations driven by noise. A long list of examples includes turbulence, critical dynamics, the dynamics of
interacting polymers, ballistic deposition (as well as other growth models) etc. The Kardar-Parisi-Zhang (KPZ)
equation that describes a growing surface under ballistic deposition \cite{kpz86, medina89} is such a model.
This equation formulated in terms of a height function $h\left( {\vec r,t} \right)$ driven by external noise is
given by
\begin{equation}
\frac{{\partial h\left( {\vec r,t} \right)}}{{\partial t}} = \nu
\nabla ^2 h + \frac{\lambda }{2}\left( {\nabla h} \right)^2  +
\eta \left( {\vec r,t} \right)
 \label{1},
\end{equation}
where $\nu$ is a diffusion constant, $\lambda$ is the coupling constant (that controls the sticking rate of the
deposited material), and $\eta \left( {\vec r,t} \right)$ is a noise term driving the equation that models the
randomness of the falling material. The noise is usually chosen to be Gaussian, with zero mean and second moment
\begin{equation}
\left\langle {\eta \left( {\vec r,t} \right)\eta \left( {\vec r',t'} \right)} \right\rangle  = 2D_0 \delta^d
\left( {\vec r - \vec r'} \right)\delta \left( {t - t'} \right)
 \label{2},
\end{equation}
where $d$ is the substrate dimension and $D_0$ specifies the noise amplitude.

The KPZ equation has been suggested 17 years ago \cite{kpz86} as an extension of the linear Edawards-Wilkinson
equation \cite{EW}, so that a lot of research has been done on the statistical properties of the surfaces that
this equation grows. An extensive review of this work can be found in refs.
\cite{barabasi95,meakin93,halpin95,krug97}. It is well known that KPZ surfaces are self-affine, and are well
described by two scaling exponents, namely the roughness exponent $\alpha$ and the dynamic exponent $z$. It
turns out that in the KPZ system these two exponents are not independent. Due to symmetry of eq. (\ref{1}) with
respect to infinitesimal tilting \cite{barabasi95} the famous scaling relation $\alpha + z = 2$ is established.
Furthermore, for the special case when $d=1$, the existence of a fluctuation-dissipation theorem gives the exact
result $\alpha  = {1 \mathord{\left/ {\vphantom {1 2}} \right. \kern-\nulldelimiterspace} 2}$ and $z = {3
\mathord{\left/ {\vphantom {3 2}} \right. \kern-\nulldelimiterspace} 2}$.

In recent years there has been a growing interest in the dynamical properties of the KPZ system. A question of
great interest regards the long-time behavior of the dynamical structure factor $\Phi _q \left( t \right) =
\left\langle {h_q \left( 0 \right)h_{ - q} \left( t \right)} \right\rangle _S $, where $h_q \left( t \right)$ is
the Fourier transform of the height function $h\left( {\vec r,t} \right)$, and $\left\langle  \cdots
\right\rangle _S $ denotes steady-state averaging over the noise. Notice that by definition $\Phi _q \left( 0
\right) = \phi _q $, where $\phi _q $ is the static two-point function.

Using a self-consistent approach, Schwartz and Edwards were able to predict a stretched exponential decay for
$\Phi _q \left( t \right)$ \cite{ES02,SE00}. Regarding the KPZ system, they found the following long-time
asymptotic behavior
\begin{equation}
\Phi _q \left( t \right) \sim c\phi _q \left( {\gamma qt^{{1 \mathord{\left/
 {\vphantom {1 z}} \right.
 \kern-\nulldelimiterspace} z}} } \right)^{\frac{{d - 1}}{2}} {\mathop{\rm e}\nolimits} ^{ - \gamma qt^{{1 \mathord{\left/
 {\vphantom {1 z}} \right.
 \kern-\nulldelimiterspace} z}} }
 \label{3},
\end{equation}
where $c$ and $\gamma $ are dimensionless constants (not necessarily $1$), $d$ is the dimension, and $z$ is the
dynamic exponent. The same asymptotic behavior was also predicted analytically by Colaiori and Moore
\cite{Moore01b} using a mode-coupling approach. Later, Colaiori and Moore solved numerically the mode-coupling
equations in one dimension \cite{Moore01c} and confirmed the asymptotic analysis for the long-time behavior.
Surprisingly, they also found that $\Phi _q \left( t \right)$ decays to zero in an oscillatory manner - a fact
that was not revealed by the analytical tools.

It should be stressed however that the above describe only
approximations to a solution of the KPZ equation. Even an exact
solution of either the Self-Consistent Expansion (SCE) equation or
the mode-coupling equation would provide only an approximation for
the real, time-dependant structure factor, $\Phi _q \left( t
\right)$, of the KPZ equation. Indeed, a more recent publication
by Pr\"ahofer and Spohn suggests that the envelope of the
one-dimensional structure factor decays exponentially rather than
as a stretched exponential \cite{spohn}. They claim an exact
solution for the time-dependant structure factor of another model
in the universality class of the KPZ system, namely the
polynuclear growth model. The actual solution for the structure
factor follows a number of well defined steps involving some
direct though complicated numerical calculations. Those
calculations are reported to be performed with extremely high
precision that seems to ensure that the final solution for the
structure factor is not affected by inaccuracies in the numerical
procedure. Although the solution they present is numerical it is
rather obvious that the envelope decay is exponential rather than
stretched exponential.

This discrepancy motivated us to check the above results directly
on the KPZ equation. In this work we find numerical support for
the existence of stretched exponential relaxations in the KPZ
system in $1 + 1$ dimensions in contradiction to ref.
\cite{spohn}. We also get direct evidence for the oscillatory
behavior of $\Phi _q \left( t \right)$. As a by product, we were
able to verify the predicted short-time behavior of the dynamical
structure factor (given in ref. \cite{Moore01a} for example), and
the validity of the scaling hypothesis for small $q$'s.

We discretized the KPZ equation (\ref{1}) on a one dimensional lattice, with lattice constant $\Delta x$, and
time difference $\Delta t$,
\begin{eqnarray}
 h\left( {x,t + \Delta t} \right) &=& h\left( {x,t} \right) +  \nonumber\\
 &+& \frac{{\Delta t}}{{\left( {\Delta x} \right)^2 }}\sum\limits_{i = 1}^d {\left\{ {\nu \left[ {h\left( {x +
\Delta x,t} \right) - 2h\left( {x,t} \right) + h\left( {x - \Delta x,t} \right)} \right]\begin{array}{*{20}c}
   {}  \\
   {}  \\
\end{array}} \right.}  \nonumber \\
 &+& \left. {  \frac{\lambda }{8}\left[ {h\left( {x + \Delta x,t} \right) - h\left( {x - \Delta x,t} \right)} \right]^2 } \right\} + \sigma \left( {12\Delta t} \right)^{{1 \mathord{\left/
 {\vphantom {1 2}} \right.
 \kern-\nulldelimiterspace} 2}} \eta \left( t \right)
\label{4},
\end{eqnarray}
where $\sigma ^2  \equiv {{2D_0 } \mathord{\left/ {\vphantom {{2D_0 } {\Delta x}}} \right.
\kern-\nulldelimiterspace} {\Delta x}}$ and the random numbers $\eta \left( t \right)$ are uniformly distributed
between $ - {1 \mathord{\left/ {\vphantom {1 2}} \right. \kern-\nulldelimiterspace} 2}$ and ${1 \mathord{\left/
{\vphantom {1 2}} \right. \kern-\nulldelimiterspace} 2}$. In this work we used $L = 1024$, $\Delta t = 0.05$,
$\Delta x = 1$, $\nu  = 1$, $\lambda  = 4$ and $D_0  = {1 \mathord{\left/ {\vphantom {1 6}} \right.
\kern-\nulldelimiterspace} 6}$. After reaching steady state, at each time step we Fourier transformed the
discrete height function, and obtained $h_q \left( t \right)$ for $q = q_0 ,2q_0 , \ldots $, where $q_0  =
{{2\pi } \mathord{\left/ {\vphantom {{2\pi } {1024}}} \right. \kern-\nulldelimiterspace} {1024}}$. Then we
calculated the two-point function $\left\langle {h_q \left( 0 \right)h_{ - q} \left( t \right)} \right\rangle _S
$ by averaging over all pairs with a time difference $t$, for $q = 30q_0 ,60q_0 ,120q_0 $ and $240q_0 $.

\begin{figure}[htb]
\includegraphics[width=6cm]{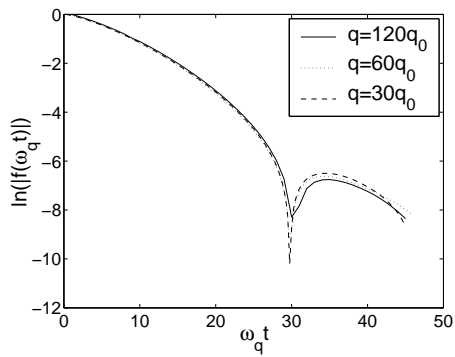}
\caption{A log plot of the scaling function $f\left( {\omega _q t} \right)$ for various small $q$'s.}
\label{Fig1}
\end{figure}

First, the scaling hypothesis was numerically verified, i.e. In Fig. \ref{Fig1} we plot $f\left( {\omega _q t}
\right) \equiv {{\Phi _q \left( t \right)} \mathord{\left/ {\vphantom {{\Phi _q \left( t \right)} {\phi _q }}}
\right. \kern-\nulldelimiterspace} {\phi _q }}$ as a function of $\omega _q t = Bq^z t$ with $z = {3
\mathord{\left/ {\vphantom {3 2}} \right. \kern-\nulldelimiterspace} 2}$ for $q = 30q_0$, $q = 60q_0$ and $q =
120q_0 $. The plot indicates good scaling.

\begin{figure}[htb]
\includegraphics[width=6cm]{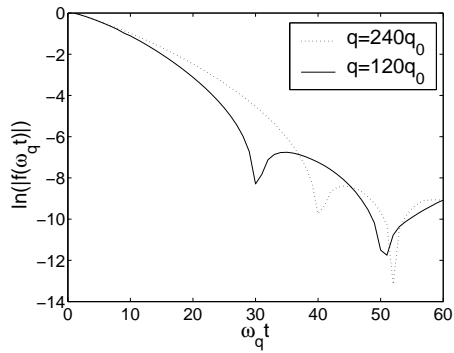}
\caption{A log plot of the scaling function $f\left( {\omega _q t} \right)$ for $q=120q_0$ and $q=120q_0$.}
\label{Fig15}
\end{figure}

However, we found that the scaling hypothesis breaks down when taking $q = 240q_0$ (see Fig. \ref{Fig15}) which
means that the scaling, which is supposed to be correct for small $q$'s, is correct only up to $q$'s that are of
order of $\sim 10\% $ of the largest $q$ in the system.

\begin{figure}[htb]
\includegraphics[width=6cm]{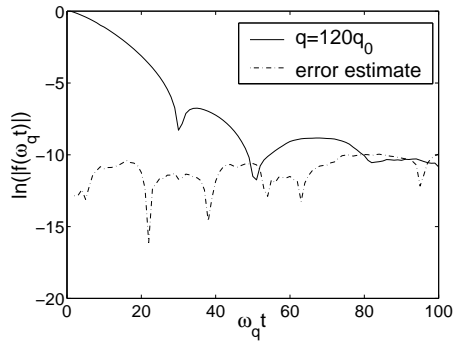}
\caption{A log plot of $f\left( {\omega _q t} \right)$ for $q = 120q_0$ with an error estimate.}
\label{Fig2}
\end{figure}

Fig. \ref{Fig1} indicates good scaling at least up to $120q_0 $. Therefore, we invested most of the
computational effort in this Fourier component, since eq. (\ref{3}) indicates that the larger $q$ we take, the
faster computational time evolution we get. For $q = 120q_0 $ we took $5 \cdot 10^{10} $ integration time steps
of $\Delta t$. Taking this component we found clear evidence for oscillatory behavior as shown in Fig.
\ref{Fig2}.

The error estimate in Fig. \ref{Fig2} was obtained using the imaginary part of $\left\langle {h_q \left( 0
\right)h_{ - q} \left( t \right)} \right\rangle _S$ after averaging. Note that $\left\langle {h_q \left( 0
\right)h_{ - q} \left( t \right)} \right\rangle _S$ should be real, due to averaging, while each contribution of
the form $h_q \left( 0 \right)h_{ - q} \left( t \right)$ is certainly not real. This means that the imaginary
part, which should vanish eventually, is a sensible error estimate. An independent argument for the estimation
of the error, which yields the same order of magnitude runs as follows: consider the case where there is no
correlation at all and estimate the "apparent correlation" due to the finite sample. The total number of time
steps is $N=5 \cdot 10^{10}$. Therefore, this is also the number of pairs $(h_q,h_{-q})$ separated by a time
$t=n\Delta t$ with $n \ll N$. The measured correlation is
\begin{equation}
\left\langle {h_q \left( 0 \right)h_{ - q} \left( t \right)} \right\rangle  = \frac{1}{N}\sum\limits_{i = 1}^N
{h_q \left( {i\Delta t} \right)h _{ - q} \left[ {\left( {i + n} \right)\Delta t} \right]}
 \label{5}.
\end{equation}
In the absence of correlation, the sum on the right hand side of (\ref{5}) is a sum of $N$ random variables.
Therefore, the size of the apparent correlation is of order $N^{ - {1 \mathord{\left/ {\vphantom {1 2}} \right.
\kern-\nulldelimiterspace} 2}}$, which is of the order of $e^{-12}$.

\begin{figure}[htb]
\includegraphics[width=6cm]{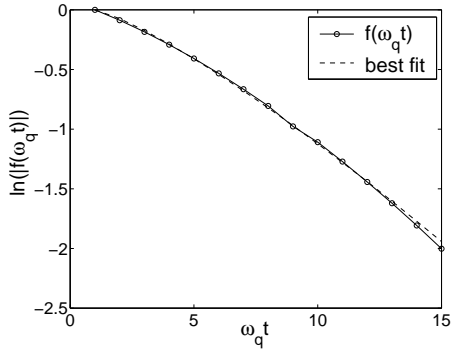}
\caption{A log plot of $f\left( {\omega _q t} \right)$ for small $\omega_qt$'s.} \label{Fig25}
\end{figure}

Using this result we extracted the short-time behavior of the scaling function. In ref. \cite{Moore01a} Colaiori
and Moore predict $\Phi _q \left( t \right) \propto \phi _q \left[ {1 - \left( {\omega _q t} \right)^{{\Gamma
\mathord{\left/ {\vphantom {\Gamma  z}} \right. \kern-\nulldelimiterspace} z}} } \right]$. Now, in one dimension
$\Gamma  = 2$ ($\Gamma $ is related to the roughness exponent $\alpha $ via $\Gamma  = d + 2\alpha $) and $z =
{3 \mathord{\left/ {\vphantom {3 2}} \right. \kern-\nulldelimiterspace} 2}$, so that for a specific $q$ we get
$f\left( {\omega _q t} \right) \propto 1 - \left( {\omega _q t} \right)^{{4 \mathord{\left/ {\vphantom {4 3}}
\right. \kern-\nulldelimiterspace} 3}}$. This prediction was indeed verified by our data as shown in Fig.
\ref{Fig25}.

\begin{figure}[htb]
\includegraphics[width=6cm]{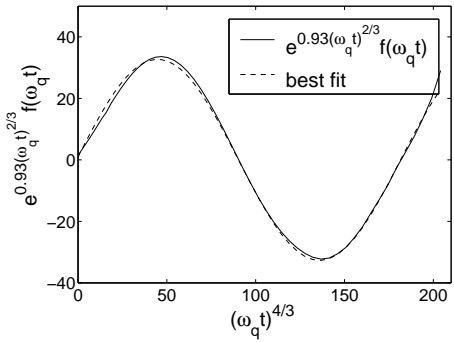}
\caption{A simple fit is $\left\{{\cos \left[ {\beta \left( {\omega _q t} \right)^{4/3}}\right]+ D\sin \left[
{\beta \left( {\omega _q t} \right)^{4/3}} \right]} \right\}$ with $\beta=0.034$ and $D=32.7$.} \label{Fig3}
\end{figure}

The error estimates suggest that the data may be useful in the range $0<\omega_qt<55$. Since we wanted to check
a stretched exponential decay, we multiplied our numerical $f(\omega_qt)$ by $\exp[\gamma\omega_qt]^{2/3}$. We
chose $\gamma=0.93$ so as to render the resulting function to be oscillating in that region. Motivated by the
predictions of ref. \cite{Moore01a} and our previous verification of it we present the resulting function as a
function of $(\omega_qt)^{4/3}$ in Fig. \ref{Fig3}.

We were thus led to try the fit ${\mathop{\rm e}\nolimits} ^{ -
\gamma \left( {\omega _q t} \right)^{{2 \mathord{\left/ {\vphantom
{2 3}} \right. \kern-\nulldelimiterspace} 3}} } \left\{ {\cos
\left[ {\beta \left( {\omega _q t} \right)^{{4 \mathord{\left/
{\vphantom {4 3}} \right. \kern-\nulldelimiterspace} 3}} } \right]
+ D\sin \left[ {\beta \left( {\omega _q t} \right)^{{4
\mathord{\left/ {\vphantom {4 3}} \right.
\kern-\nulldelimiterspace} 3}} } \right]} \right\}$ for the full
scaling function $f(\omega_qt)$. This has the right small
$\omega_qt$ behavior as well as the envelope stretched exponential
decay.

\begin{figure}[htb]
\includegraphics[width=6cm]{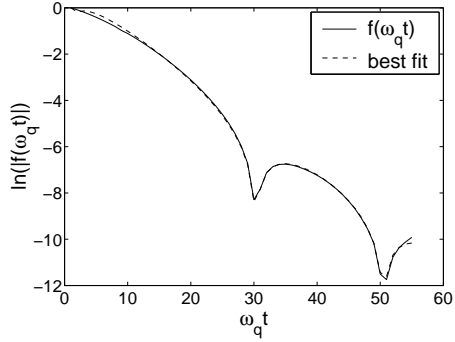}
\caption{A fit of $f\left( {\omega _q t} \right)$ using $e ^{ - \gamma (\omega _q t)^{2/3}} \left\{{\cos \left[
{\beta \left( {\omega _q t} \right)^{4/3}}\right]+ D\sin \left[ {\beta \left( {\omega _q t} \right)^{4/3}}
\right]} \right\}$ with $\gamma=0.93$, $\beta=0.034$ and $D=32.7$. Both curves practically join over most of the
time range.} \label{Fig4}
\end{figure}

\begin{figure}[htb]
\includegraphics[width=6cm]{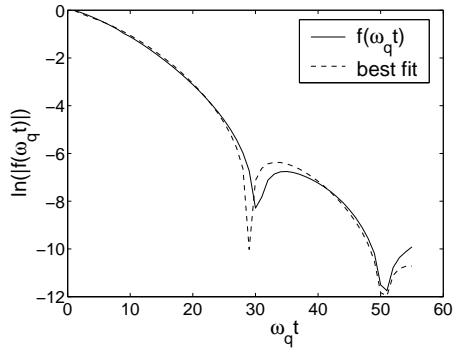}
\caption{A fit of $f\left( {\omega _q t} \right)$ using $e ^{ - \gamma (\omega _q t)} \left\{{\cos \left[ {\beta
\left( {\omega _q t} \right)^{4/3}}\right]+ D\sin \left[ {\beta \left( {\omega _q t} \right)^{4/3}} \right]}
\right\}$. The best fit of that form is obtained for $\gamma=0.21$, $\beta=0.326$ and $D=2.6$.}
\label{Fig5}
\end{figure}

The fit is very good, especially when keeping in mind that this
fit involves more than $6$ orders of magnitude in the size of the
scaling function. An attempt to replace the ${2 \mathord{\left/
{\vphantom {2 3}} \right. \kern-\nulldelimiterspace} 3}$ in the
stretched exponential (in the ansatz) with an arbitrary exponent
that will be determined by the fitting procedure yields a very
close value ($\sim 0.65$). Actually, we also tried to fit an
exponential decay (rather than a stretched exponential one),
according to the finding of ref. \cite{spohn}, that gave a fit
that was definitely worse (Fig. \ref{Fig5}).

At this point we are facing an interesting contradiction. First, we have results following from four independent
approaches: (1) Analytical asymptotic study of the self-consistent approximation. (2) Analytical asymptotic
study of the mode-coupling approximation. (3) Numerical solution of the mode-coupling equations. (4) The present
direct numerical integration of the KPZ equation. All these independent approaches yield the same stretched
exponential decay. On the other hand, we have a publication \cite{spohn} claiming to be exact that yields a
result that is quite different from the results obtained from all the above methods. It is possible that there
is a flaw in the study presented in ref. \cite{spohn}. If there is, we have certainly not found any. We would
like to suggest here another solution to the problem. Ref. \cite{spohn} does not consider directly the KPZ
equation but rather the polynuclear growth model that was shown to be equivalent to the directed polymer problem
with specific boundary conditions, and thus in the same universality class of the KPZ system. However, the point
is that two models that are in the same universality class must have the same exponents but not necessarily the
same scaling functions. In fact, it is possible to construct families of exactly solvable models where all the
members of the family are characterized by the same exponents yet have radically different time dependant
structure factors. We will not go into this here but the interested reader could find the relevant ideas,
although presented in a different context, in ref. \cite{S03}. Our suggestion for solving the puzzle is thus
that Pr\"ahofer and Spohn \cite{spohn} obtain the correct decay for the polynuclear growth model, which is in
itself a most impressive feat, but this is not the decay of the KPZ structure factor.

To summarize, using extensive numerical integration of the KPZ
equation in $1 + 1$ dimensions, this work gives clear support for
the scaling hypothesis (i.e. the fact that the scaling function
$f\left( {\omega _q t} \right)$ is the same for any $q$, at least
for small $q$'s), verifies the short-time behavior given in ref.
\cite{Moore01a} (i.e. $\Phi _q \left( t \right) \propto \phi _q
\left[ {1 - \left( {\omega _q t} \right)^{{\Gamma  \mathord{\left/
{\vphantom {\Gamma  z}} \right. \kern-\nulldelimiterspace} z}} }
\right]$) and establishes the oscillatory decay of the dynamical
structure factor to zero (as suggested in \cite{Moore01c}). In
addition, we show that the stretched exponential describes the
decay of the structure factor over six orders of magnitude in its
size. This implies that the KPZ problem is likely to be a
respected member of the family of systems that exhibit slow
relaxations - thus opening the door for mutual influence between
the community of surface growth and that of slow dynamics.

The present work consumed quite a few months of CPU time on a Pentium 4 machine. In spite of that, we believe
that the results presented here should motivate a much heavier numerical effort to deal with a region with
larger $\omega_qt$'s for the one-dimensional system and hopefully with the two dimensional system.

\end{document}